\pdfoutput=1

\documentclass{article}

\usepackage[eandd, final]{neurips_2026}  

\usepackage[utf8]{inputenc}
\usepackage[T1]{fontenc}
\usepackage{hyperref}
\hypersetup{pdftitle={HackerSignal: A Multi-Source, CVE-Linked Corpus and Prospective Benchmark for Cyber Threat Intelligence}, pdfauthor={Benjamin M. Ampel, Sagar Samtani}}
\usepackage{url}
\usepackage{booktabs}
\usepackage{amsfonts}
\usepackage{amsmath}
\usepackage{nicefrac}
\usepackage{microtype}
\usepackage{xcolor}
\usepackage{graphicx}
\usepackage{multirow}
\usepackage{array}
\usepackage{subcaption}
\usepackage{pifont}
\usepackage{wrapfig}

\newcommand{\etg}{\textsc{HackerSignal}}

\newcommand{\cmark}{\ding{51}}
\newcommand{\xmark}{\ding{55}}

\title{\etg: A Multi-Source, CVE-Linked Corpus and Prospective
       Benchmark for Cyber Threat Intelligence}

\author{%
  Benjamin M.\ Ampel \\
  Department of Computer Information Systems \\
  J.\ Mack Robinson College of Business, Georgia State University \\
  \texttt{bampel@gsu.edu} \\
  \And
  Sagar Samtani \\
  Department of Operations and Decision Technologies \\
  Data Science and Artificial Intelligence Lab \\
  Kelley School of Business, Indiana University \\
  \texttt{ssamtani@iu.edu} \\
}

\begin{document}

\maketitle

\begin{abstract}

Cyber threat intelligence (CTI) systems must link exploit code, advisories, and fix commits to the vulnerabilities they concern, and they must do so for vulnerabilities disclosed after training. HackerSignal is a public corpus and benchmark built for this prospective setting. It aggregates \textbf{7.45 million exact-deduplicated documents} from \textbf{64 public sources} across eight source layers (hacker forums, exploit archives, exploit question-and-answer sites, vulnerability references, advisories, bug-bounty disclosures, fix commits, and known-exploited-vulnerability records), spanning 1988--2026. All layers are keyed to a shared CVE identifier space with source provenance and release-mode governance. We define three benchmark tasks with temporal splits: (1)~\emph{CVE linkage retrieval} over 340,536 National Vulnerability Database descriptions, (2)~\emph{exploit type classification} into eight classes, and (3)~\emph{temporal generalization}, where training and test CVE sets do not overlap ($C_{\text{train}} \cap C_{\text{test}} = \emptyset$). Zero-shot dense and hybrid retrievers reach a mean reciprocal rank (MRR) of 0.71 on Task~1, and the top models are statistically tied under paired bootstrap tests. Cybersecurity-pretrained encoders lead exploit type classification (macro-F1 0.90). A two-annotator audit of 240 linkage labels finds 79.5\% precision (Cohen's $\kappa$ 0.94), with per-source precision from 20\% to 100\%.
The hacker-community layer contributes 6.97 million posts but only about 2,000 CVE-linked records. A masked forum-post grounding pilot over these records reaches MRR 0.08, about nine times lower than the structured-source tasks. Grounding informal hacker discourse to vulnerabilities therefore remains an open problem that the release is designed to support. Code, data, and Croissant metadata are available at \href{https://huggingface.co/datasets/BenAmpel/HackerSignal}{\texttt{hf.co/datasets/BenAmpel/HackerSignal}} (data) and \href{https://github.com/BenAmpel/hackersignal}{\texttt{github.com/BenAmpel/hackersignal}} (code).
\end{abstract}

\section{Introduction}

Understanding how cyber threats emerge, propagate, and relate to vulnerabilities is a central challenge in modern applied cybersecurity. Signals that track this trajectory are distributed across heterogeneous text-based sources, including: (1) hacker forums where exploits are first discussed, (2) public exploit databases where proof-of-concept code is released, (3) official vulnerability advisories such as the National Vulnerability Database, CISA's Known Exploited Vulnerabilities, and GitHub Advisory where threats are cataloged, and (4) open-source repositories where fixes are committed. Extant public datasets in this domain often span a single or subset of these layers~\citep{crimebb2018} and are usually not publicly released~\citep{ampel2020dtl}. A survey of CTI extraction from text identifies advisory records and threat reports as the dominant labeled sources studied, with exploit-archive and PoC-focused benchmarks remaining scarce~\citep{rahman2023cti}; a companion survey of cybercrime community research corroborates that structured datasets linking forum posts to formal vulnerability records are largely absent from the public literature~\citep{hughes2024cybercrime}. HackerSignal addresses this gap with a temporally structured, multi-source corpus and benchmark spanning the full advisory-to-exploit trajectory, enabling downstream CTI analytics tasks. HackerSignal makes three key contributions:

\begin{enumerate}
  \item \textbf{A temporally grounded CTI corpus} of 7.45M documents from 64 public forum/source identifiers across eight source layers (hacker communities, exploit archives, vulnerability references, exploit Q\&A references, advisories, bug-bounty disclosures, fix commits, and exploitation references), with CVE linkage metadata connecting vulnerability identifiers to the texts that discuss them. The 6.97M-row hacker community subcorpus alone (spanning 56 forums across darkweb marketplaces, security discussion boards, and multilingual hacker communities) supports diverse NLP tasks beyond our three benchmarks, including exploit code summarization, security content classification, topic modeling, community interaction analysis, and cross-lingual CTI extraction.

  \item \textbf{Three benchmark tasks} with temporal train/val/test splits that impose progressively stricter constraints on training: (1) cross-source CVE linkage retrieval, (2) 8-class exploit type classification, and (3) CVE-disjoint temporal generalization ($C_{\text{train}} \cap C_{\text{test}} = \emptyset$). The Task~3 test queries are a subset of the Task~1 test queries, so the tasks differ in what training may see rather than in evaluation difficulty. Each task ships with shortcut and leakage diagnostics.

  \item \textbf{A full model ladder and governance artifacts} from BM25 through four dense bi-encoders (MiniLM, mpnet, BGE, E5), bag-of-words classifiers (Decision Tree, TF-IDF+LR, SVM), recurrent neural networks (RNN, GRU, LSTM, BiLSTM), three security-domain transformers (SecBERT, SecureBERT, CySecBERT), and hybrid sparse+dense retrieval, alongside source-shortcut diagnostics, manual-audit packets, and a release-governance addendum.
\end{enumerate}

\noindent Figure~\ref{fig:pipeline} provides an overview of the complete pipeline of the proposed dataset, including eight source layers unified through a shared CVE identifier space, then channeled into three benchmark tasks with temporal controls. Temporal splits are necessary in our context as real CTI systems must generalize to threats that did not exist at training time~\citep{koh2021wilds}. We explicitly audit the dataset for temporal and source leakage~\citep{kapoor2023leakage} because benchmark claims are otherwise easy to inflate (discussed in Section 3).

\section{Related Work}

\paragraph{Hacker-community cyber threat intelligence.} Early hacker-forum CTI work treated underground communities as a source of proactive intelligence about tools, actors, and emerging attack capabilities~\citep{samtani2015assets, samtani2017jmis}. The AZSecure Hacker Assets Portal operationalized this as a situational awareness platform~\citep{samtani2021hap}, while parallel work modeled threat emergence over time via diachronic graph embeddings~\citep{samtani2020dgef} and mobile malware identification~\citep{grisham2017mobile}. The resulting artifacts are operational systems rather than reusable ML benchmarks, since their data remains proprietary. \citet{rahman2023cti} survey automated CTI extraction from text, observing that collections often concentrate on the advisory and threat-report tier, while exploit-archive and PoC-focused benchmarks remain scarce. \citet{hughes2024cybercrime} systematize research on the cybercrime community and demonstrate a persistent shortage of ground-truth datasets. Most publicly collected forum data lacks structured labels linking posts to formal vulnerability records.

\paragraph{Exploit labeling and vulnerability linkage.}
The closest predecessor to HackerSignal is in the field of exploit labeling research. Prior methods labeled hacker-forum exploit source code via deep transfer learning from professional exploit repositories~\citep{ampel2020dtl} and later extended this with pre-initialization, multi-layer transfer, and self-attention~\citep{ampel2024misq}. For cross-source linkage, an attention-based deep-structured semantic model linked dark-web exploits to known vulnerabilities, yielding 52,590 linkages~\citep{samtani2022eva}. HackerSignal differs by releasing a governed, temporally split public benchmark that exposes source provenance, weak-label caveats, and a quality-controlled CVE-linkage retrieval task. Finally, the hacker forum collection of~\citet{otto2021isi} aggregates 259K posts from three English-language security communities (AntiOnline, Go4Expert, CipherSpace) and studies exploit-sharing behavior via graph embeddings.

\paragraph{Underground community datasets.} Research has also focused on the general crawling of hacker communities, like a credentialized dataset of darkweb cybercrime forum posts widely used for crime and exploit prediction~\citep{crimebb2018} \citet{nunes2016darknet} presented an operational darknet/deepnet mining system over hacker forums and markets for proactive CTI. The Gayanku~\citep{gayanku2020} dataset contains 3.8M posts from seven darknet marketplaces (Silk Road, Agora, Evolution, etc.) on Kaggle.

\paragraph{Exploit and vulnerability databases.} ExploitDB~\citep{exploitdb} is the salient public exploit archive (50K+ entries, many with CVE IDs). The CVEfixes dataset~\citep{cvefixes2021} links CVEs to fix commits in open-source repositories. NVD~\citep{nvd}, CISA KEV~\citep{cisakev}, and GitHub Advisory~\citep{githubadvisory} are official advisory sources. HackerSignal normalizes all five into a common JSONL schema with CVE keys.

\paragraph{Vulnerability intelligence and CTI benchmarks.} VulZoo aggregates 17 vulnerability intelligence feeds, four of them security mailing lists, into one dataset~\citep{vulzoo2024}. Its discussion layer is professional disclosure rather than underground community forums, and it provides no prospective temporal splits or benchmark tasks. CTIBench evaluates large language models on four CTI tasks, two of them CVE-grounded (CVE-to-CWE mapping and CVSS prediction)~\citep{ctibench2024}. It has no discourse layer and no temporal splits. Neither pairs community discourse with CVE-linked, temporally split benchmarks. HackerSignal combines both and adds a governed public release.

\paragraph{Multilingual CTI.} More general hacker-forum text classification work compared SVM and CNN models for extracting CTI from forum posts~\citep{deliu2017cti}, while \citet{ebrahimi2020crosslingual} showed that non-English hacker forums require cross-lingual approaches that avoid brittle machine translation. CVE severity prediction~\citep{severitypred2017}, exploit labeling~\citep{ampel2024misq}, and exploit-vulnerability linkage~\citep{samtani2022eva} all motivate our task design; HackerSignal operationalizes them as public, temporally split benchmarks.

\paragraph{Benchmark design and evaluation method.} BEIR~\citep{thakur2021beir} established that retrieval benchmarks require heterogeneous baseline ladders rather than single-model comparisons. KILT~\citep{petroni2021kilt} showed that knowledge-intensive linking tasks should reward both task success and provenance quality. WILDS~\citep{koh2021wilds} formalized natural arising distribution shifts (including temporal shifts) as first-class benchmark objects. At the same time, WRENCH~\citep{zhang2021wrench} demonstrated that weak-supervision evaluation is sensitive to hidden protocol choices, requiring explicit reporting of label sources and aggregation rules. \citet{kapoor2023leakage} documented widespread data leakage in ML-for-science benchmarks and proposed structured reporting to make integrity checks legible. These methodological precedents directly inform our temporal splits, source-shortcut diagnostics, and audit artifacts.

\section{Dataset}
\label{sec:dataset}

\subsection{Source Layers}

\begin{figure}[t]
\centering
\includegraphics[width=0.8\linewidth]{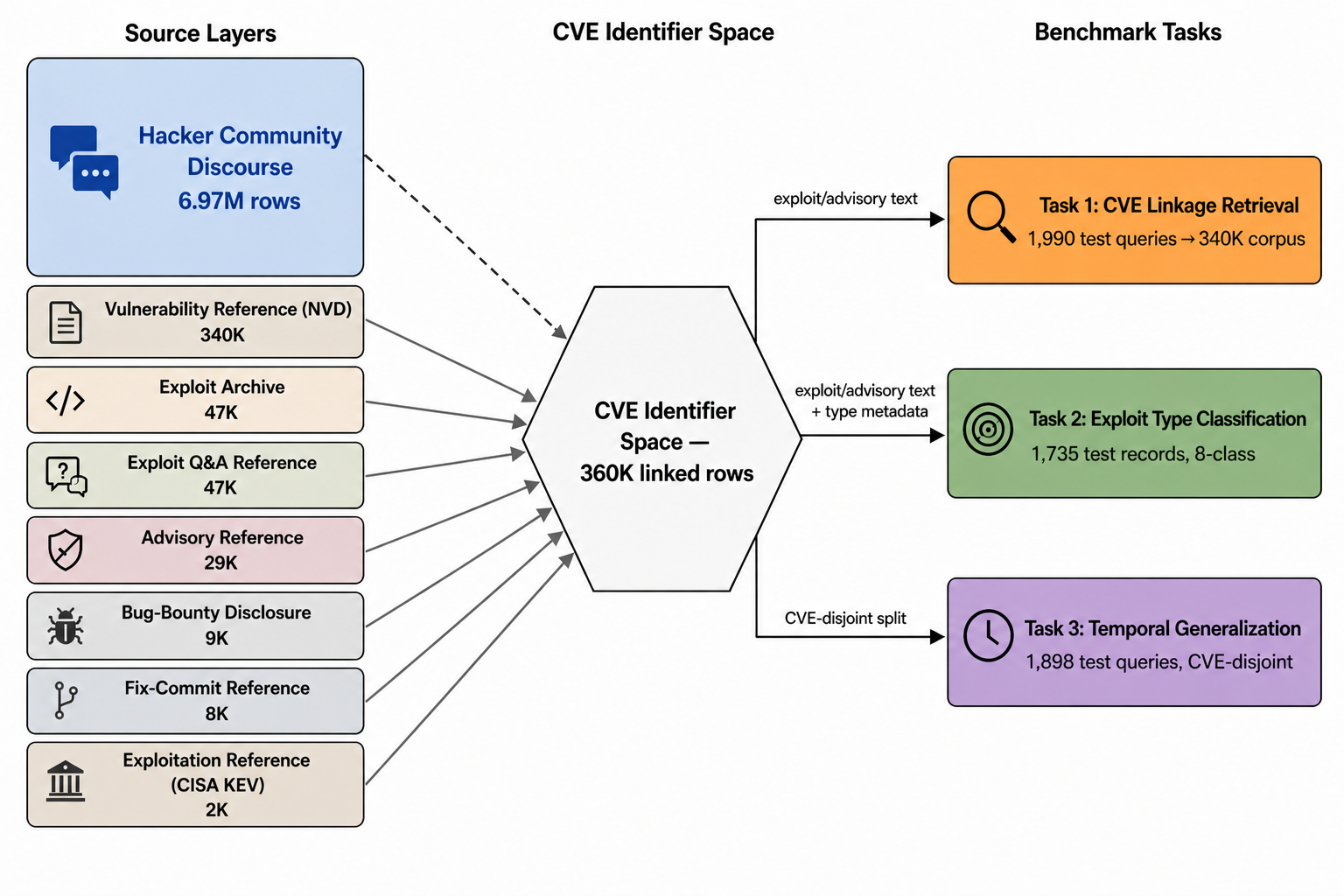}
\caption{HackerSignal pipeline overview. Dashed arrows indicate sparse, indirect CVE mentions. Solid arrows indicate direct CVE metadata linkage.}
\label{fig:pipeline}
\end{figure}

HackerSignal integrates public hacker-community discourse with sources of exploit, advisory, vulnerability-reference, bug-bounty, and commit-fix in eight source layers (Figure~\ref{fig:pipeline}, left). Table~\ref{tab:corpus-stats} reports the composition of the audited release; Table~\ref{tab:quality-audit} reports quality diagnostics.

\begin{table}[t]
\centering
\small
\setlength{\tabcolsep}{3pt}
\caption{Dataset statistics after exact deduplication. The hacker community layer is expanded by forum. The remaining layers are single-source. Avg.\ Len = mean whitespace tokens; CVE\% = rows containing an explicit CVE identifier.}
\label{tab:corpus-stats}
\begin{tabular}{llrrrl}
\toprule
\textbf{Source Layer} & \textbf{Source / Forum} & \textbf{\#Docs} & \textbf{Avg Len} & \textbf{CVE\%} & \textbf{Span} \\
\midrule
\multirow{9}{*}{\rotatebox[origin=c]{90}{\footnotesize Hacker Community}}
 & gayanku\_darkweb     & 3,107,038 & 72.6 & $<$0.1 & 2011--17 \\
 & antichat             & 1,819,730 & 39.7 & $<$0.1 & 2002--26 \\
 & evolution\_darkweb   &   471,115 & 65.4 & $<$0.1 & 2014--15 \\
 & antionline           &   462,510 & 73.6 & $<$0.1 & 2001--26 \\
 & persiantools         &   280,293 & 82.1 & $<$0.1 & 2026 \\
 & wwhclub              &   161,026 & 51.3 & $<$0.1 & 2026 \\
 & hackforums           &   151,072 & 31.6 & $<$0.1 & 2009--15 \\
 & go4expert            &   110,345 & 75.1 & $<$0.1 & 2004--26 \\
 & \textit{Other (48 forums)} &   402,885 & 56.5 & 0.1 & 1996--26 \\
\cmidrule{2-6}
 & \textbf{Subtotal}    & \textbf{6,966,014} & \textbf{61.7} & \textbf{$<$0.1} & \textbf{1996--26} \\
\midrule
\multicolumn{2}{l}{Vulnerability Ref (NVD)} &   340,536 & 47.2 & 100.0 & 2002--26 \\
\multicolumn{2}{l}{Exploit Archive}         &    46,653 & 240.0 & 10.5 & 1988--26 \\
\multicolumn{2}{l}{Exploit Q\&A Ref}        &    46,505 & 25.2 & $<$0.1 & 2000 \\
\multicolumn{2}{l}{Advisory Ref}            &    28,684 & 110.8 & 5.0 & 2017--26 \\
\multicolumn{2}{l}{Bug-Bounty Disclosure}   &     9,453 & 191.1 & 6.5 & 2013--24 \\
\multicolumn{2}{l}{Fix-Commit Ref}          &     8,293 & 92.0 & 100.0 & 1999--22 \\
\multicolumn{2}{l}{Exploitation Ref (KEV)}  &     1,508 & 54.4 & 3.8 & 2021--26 \\
\midrule
\multicolumn{2}{l}{\textbf{Total (64 sources)}} & \textbf{7,447,646} & \textbf{62.0} & \textbf{4.8} & \textbf{1988--26} \\
\bottomrule
\end{tabular}
\end{table}

\begin{table}[t]
\vspace{-4pt}
\centering
\small
\setlength{\tabcolsep}{4pt}
\caption{Corpus quality diagnostics.}
\label{tab:quality-audit}
\begin{tabular}{lrr}
\toprule
\textbf{Diagnostic} & \textbf{Rows} & \textbf{Share} \\
\midrule
Parseable timestamps        & 7,447,646 & 100.0\% \\
CTI-keyword nonzero         & 1,748,049 &  23.5\% \\
CVE string present          &   360,004 &   4.8\% \\
Short rows ($<$8 tok)       & 1,569,263 &  21.1\% \\
Long rows ($>$3000 tok)     &       815 &  $<$0.1\% \\
Est.\ near-duplicates       &   609,029 &   8.2\% \\
\bottomrule
\end{tabular}
\vspace{-4pt}
\end{table}

\paragraph{Hacker and security communities (6.97M rows).} The hacker-community layer captures organic security discourse. This includes, but is not limited to, discussions of exploits, tool sharing, vulnerability commentary, marketplace posts, and peer support across communities. Our collection combines historical forum datasets (e.g., Gayanku, Evolution, and HackForums) with newer publicly accessible sources such as DeepDarkCTI communities, 0x00sec, HackerSploit, and ParrotSec. The largest forums after deduplication are \texttt{gayanku\_darkweb} (3.11M rows), \texttt{antichat} (1.82M), \texttt{evolution\_darkweb} (471K), \texttt{antionline} (463K), \texttt{persiantools} (280K), \texttt{wwhclub} (161K), and \texttt{hackforums} (151K). Forum sources often carry sparse direct CVE mentions and exhibit heterogeneous vocabulary overlap across communities (Figure~\ref{fig:domain-overlap}). These characteristics make CVE linkage from community discourse a nontrivial retrieval problem.

\paragraph{Reference and supervision layers (482K rows).}
The remaining layers provide CVE-grounded supervision and comparison corpora. NVD CVE descriptions (340,536 rows) form the retrieval corpus for all three benchmark tasks. ExploitDB, 0day.today, Packet Storm, ZeroScience, and Vulnerability-Lab contribute exploit and advisory text. GitHub Advisory, CISA KEV, HackerOne, and CVEfixes contribute structured records for vulnerabilities, bug bounties, known-exploited issues, and fix commits. These sources are separated by \texttt{source\_layer}; thus, experiments can evaluate either community-only discourse or community-plus-context settings.

\subsection{CVE Identifier Space}

All sources are unified through a common JSONL record format and a shared CVE identifier namespace (Figure~\ref{fig:pipeline}, center). Each record carries a SHA-256-derived release identifier, source provenance (\texttt{source\_layer}, \texttt{forum\_id}), cleaned UTF-8 text, an ISO~8601 timestamp, and a pseudonymized author hash. A separate CVE index maps each record to its CVE reference(s) via a foreign key, enabling cross-source joins through the shared identifier space. Of the 7.45M documents, 360,004 rows (4.83\%) contain explicit CVE references, forming the linkage substrate for all three benchmark tasks. The full schema specification is provided in the accompanying datasheet~\citep{gebru2021datasheet}.

\subsection{Collection and Preprocessing}

Data was collected between January and April 2026 via (i)~official APIs (NVD, GitHub Advisory, CISA KEV, HackerOne HuggingFace Hub), (ii)~ingestion of existing public datasets (e.g., Gayanku, HackForums, CVEfixes, DeepDarkCTI, CrackingArena), and (iii)~ethical web scraping of publicly indexed sites (e.g., ExploitDB, Vulnerability-Lab, ZeroScience, 0x00sec, HackerSploit, ParrotSec, Hack The Box, Go4Expert, Full Disclosure). All collection code is released alongside the dataset.

\begin{figure}[t]
\centering
\includegraphics[width=\linewidth]{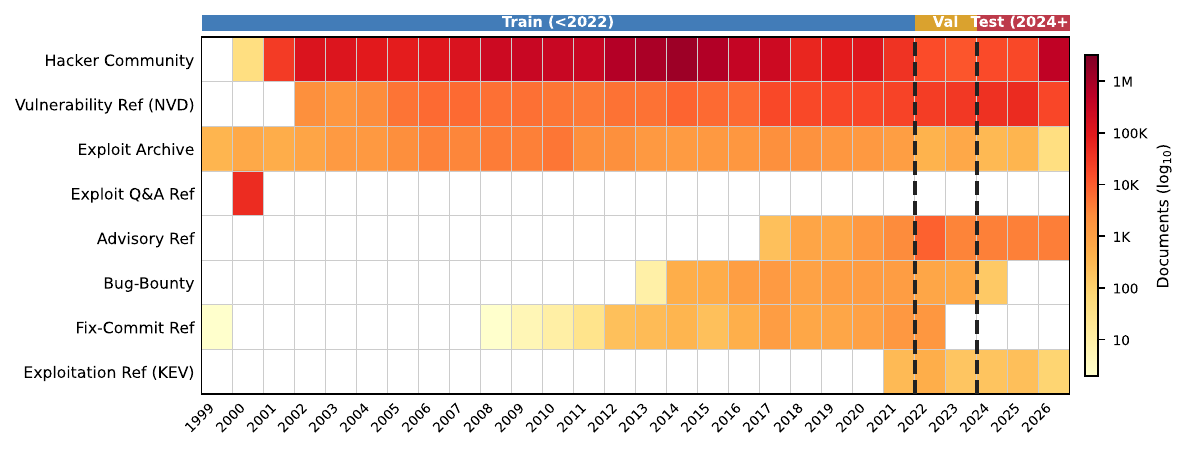}
\caption{Temporal coverage by source layer (log$_{10}$ scale). Colored bars indicate benchmark split boundaries: train ($<$2022), val (2022--23), test (2024+).}
\label{fig:temporal-heatmap}
\end{figure}

We took three steps to prepare the data. First, we clean text by normalizing Unicode, stripping HTML tags, replacing URLs, and collapsing extra whitespace. Second, we remove exact duplicates by SHA-256 hashing of the normalized text, which drops 814,922 of 8,262,749 input rows (plus 181 rows with empty text). We estimate near-duplicates with 64-bit SimHash over 5-token shingles on a deterministic 3\% sample (Hamming distance $\leq$3; Table~\ref{tab:quality-audit}) but retain them, so users can apply their own threshold. Third, we divide the posts into temporal train, validation, and test splits. We retain non-English and short posts. Non-English content is a released property of the corpus: of the 1,964 CVE-linked hacker-community records, 594 (30.2\%) are Cyrillic-heavy, and antichat (1.82M rows) and persiantools (280K) are among the eight largest forums. We show the final dataset distribution in Figure~\ref{fig:temporal-heatmap}.

Several high-volume historical sources (HackForums, Gayanku, Evolution, and the AntiOnline/Go4Expert forums) end in 2015-2017, while newer sources (DeepDarkCTI communities, 0x00sec, HackerSploit, ParrotSec, NVD, GitHub Advisory, and CISA KEV) extend into 2026 (Figure~\ref{fig:temporal-heatmap}). Non-English content (Russian, Turkish, Persian, Arabic, Chinese) appears in antichat, persiantools, DeepDarkCTI, and several dark-market sources. Short rows ($<$8 tokens, 21.1\%) include replies and marketplace acknowledgments. These should be filtered for classification and retrieval tasks, but are useful for other analyses of community interactions.

\begin{figure}[t]
\centering
\includegraphics[width=0.8\linewidth]{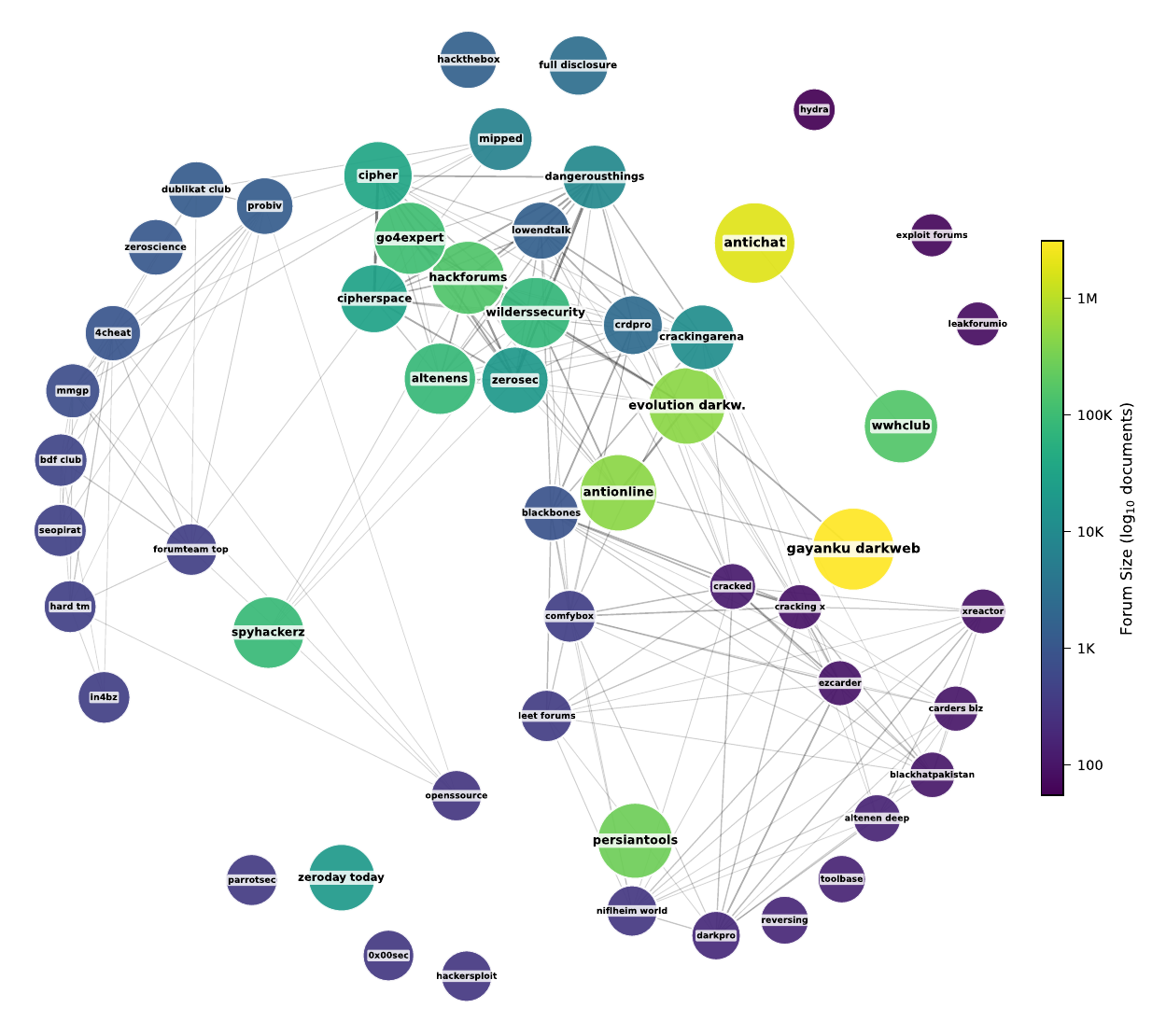}
\caption{Pairwise vocabulary overlap across 46 hacker community forums (5\% deterministic token sample, hierarchically clustered).}
\label{fig:domain-overlap}
\end{figure}

Past research indicates that hacker community platforms vary in the content they offer, such as exploits and tools. To measure lexical similarity among 46 forums, we compute pairwise Jaccard coefficients using a 5\% token sample and perform hierarchical clustering (Figure~\ref{fig:domain-overlap}). The heatmap and network show a dense cluster of English-language security forums (hackforums, zerosec, cipher, wilderssecurity, go4expert, cipherspace) with shared technical vocabulary. Dark Web marketplaces (Gayanku, Evolution) form a second cluster connected by transactional terms. Non-English forums (antichat, persiantools) and niche groups (parrotsec, zeroday~today) are peripheral with low overlap ($J < 0.06$), showing language and domain create vocabulary boundaries. Node sizes reflect contribution on a $log_{10}$ scale, highlighting antichat, gayanku, and wwhclub's prominence in document count despite lexical isolation. This heterogeneity supports our evaluation, as models must generalize across diverse vocabularies.

\section{Benchmark Tasks and Results}
\label{sec:tasks}

HackerSignal provides a unique longitudinal, multi-lingual, and multi-sourced CTI dataset that can enable novel research opportunities examining temporal phenomena, generalization, and more. We illustrate three sample CTI tasks with increasing levels of methodological complexity that HackerSignal enables in (Figure~\ref{fig:pipeline}, right; Table~\ref{tab:task-summary}). Each probes a different facet of temporal out-of-distribution generalization. All tasks use temporal splits, ensuring that test data postdate the training data. Following~\citet{thakur2021beir}, we evaluate each task with a heterogeneous baseline ladder spanning lexical, dense, and hybrid methods.

\begin{table}[t]
\centering
\small
\caption{Benchmark task summary. Task~3 splits by CVE year (test 2024+); Task~1 splits by source date, so 92 of its test queries link to pre-2024 CVEs. Task~2: train $<$2020, test $\geq$2023.}
\label{tab:task-summary}
\begin{tabular}{llrrrr}
\toprule
\textbf{Task} & \textbf{Type} & \textbf{Train} & \textbf{Val} & \textbf{Test} & \textbf{Corpus / Total} \\
\midrule
1. CVE-R & Retrieval & 56,692 & 2,584 & 1,990 & 340,536 \\
2. ETC & 8-class classif. & 64,413 & 4,735 & 1,735 & 70,883 \\
3. TG & CVE-disjoint Retr. & 56,833 & 2,535 & 1,898 & 340,536$^\dagger$ \\
\bottomrule
\multicolumn{6}{l}{\footnotesize $^\dagger$Same corpus; zero CVE overlap between Task~3 train/val/test gold sets.}
\end{tabular}
\end{table}

All benchmark labels are derived from platform-provided and standardized metadata rather than manual annotation (helping reduce third-party biases). Following~\citet{zhang2021wrench}, we explicitly report weak-label sources and aggregation rules, as minor protocol changes can alter benchmark results and conclusions. For retrieval tasks (Tasks~1 and~3), a query's gold CVE is the metadata-linked entry from its source. This may be incomplete or linked to a related rather than maximally specific CVE. For exploit type classification (Task~2), labels are derived from structured metadata fields (ExploitDB exploit type and description, exploit type, HackerOne vulnerability type prefix) and mapped to a unified 8-class taxonomy via deterministic rules. We provide a 240-row stratified manual audit packet in \texttt{data/audits/} for transparency.

\paragraph{Label quality.} Two annotators independently coded the 240 audit rows. Agreement on CVE linkage is 0.971 raw (Cohen's $\kappa$ 0.942, Gwet's AC1 0.942), and agreement on exploit relevance is 0.963 raw ($\kappa$ 0.919, AC1 0.930); Gwet's AC1 is a chance-corrected coefficient that stays stable under skewed labels. Audited linkage precision is 182 of 229 adjudicated rows (79.5\%). Eleven rows were not adjudicable, so precision over the full sample lies between 75.8\% and 80.4\%. The relevance audit confirms the benchmark label in 174 of 231 rows (75.3\%). Fifty rows (20.8\%) could defensibly match more than one CVE, and 48 (20.0\%) describe more than one vulnerability. A model that retrieves a valid secondary CVE is scored incorrect, so reported recall and MRR are conservative. Reliability varies sharply by source (Table~\ref{tab:source-reliability}): feeds that republish NVD advisories carry verbatim overlap with their targets, and ZeroScience carries none. We therefore mark ZeroScience and HackerOne labels as unsuitable for supervision without further filtering.

\begin{table}[t]
\centering
\small
\caption{Per-source label reliability from the 240-row audit. Precision = correct CVE links over adjudicated rows. Overlap = rows sharing a verbatim span with the gold NVD description.}
\label{tab:source-reliability}
\begin{tabular}{lrr}
\toprule
\textbf{Source} & \textbf{Precision} & \textbf{Overlap} \\
\midrule
CVEfixes          & 28/28 (100\%)  & 28/28 \\
GitHub Advisory   & 42/42 (100\%)  & 39/43 \\
Vulnerability-Lab & 14/14 (100\%)  & -- \\
0day.today        & 11/11 (100\%)  & -- \\
DTL-EL archives   & 13/14 (92.9\%) & -- \\
CISA KEV          & 38/41 (92.7\%) & 26/43 \\
HackerOne         & 28/39 (71.8\%) & 11/42 \\
ZeroScience       & 8/40 (20.0\%)  & 0/42 \\
\midrule
All               & 182/229 (79.5\%) & 112/240 \\
\bottomrule
\end{tabular}
\end{table}

\subsection{Task 1: CVE Linkage Retrieval (CVE-R)}
\label{sec:task1}

Let $\mathcal{Q}$ be a set of exploit/advisory evidence texts and $\mathcal{C} = \{c_1, \ldots, c_N\}$ the NVD corpus of $N{=}340{,}536$ CVE descriptions. For each query $q \in \mathcal{Q}$, the task is to retrieve the correct CVE entry $c^* \in \mathcal{C}$ such that the source metadata links $q$ to $c^*$. This is cross-source temporal out-of-domain entity grounding, akin to knowledge-intensive entity linking~\citep{petroni2021kilt}, except the target knowledge base is the CVE ecosystem. Queries come from exploit archives, advisories, fix commits, and bug bounty reports, while the target space is a single, canonical CVE registry. We retain queries from 22 source identifiers spanning exploit archives, advisories, fix commits, and bug bounty reports (see \texttt{data/benchmark\_v2/task1\_cve\_linkage/meta.json} for the full list) and require $\geq$8 query tokens. Labels are metadata-derived linkages, not manually adjudicated ground truth. 

\begin{wraptable}{r}{0.52\textwidth}
\vspace{-4pt}
\centering
\small
\setlength{\tabcolsep}{3pt}
\caption{Task 1 (CVE-R) test results (1,990 queries, 340K corpus). Bootstrap 95\% intervals on MRR@10 (10,000 resamples) have a half-width of about 0.02; the top two MRR values do not differ significantly.}
\label{tab:task1-results}
\begin{tabular}{lcccc}
\toprule
\textbf{Model} & \textbf{R@1} & \textbf{R@5} & \textbf{R@10} & \textbf{MRR} \\
\midrule
BM25 & 0.595 & 0.688 & 0.715 & 0.636 \\
MiniLM-L6-v2 & 0.655 & 0.776 & 0.809 & 0.709 \\
mpnet-base-v2 & 0.626 & 0.756 & 0.792 & 0.682 \\
BGE-base-v1.5 & 0.653 & \textbf{0.789} & 0.823 & \textbf{0.714} \\
E5-base-v2 & 0.649 & \textbf{0.789} & \textbf{0.824} & 0.710 \\
\midrule
Hybrid BM25+mpnet & \textbf{0.679} & 0.753 & 0.768 & 0.711 \\
\bottomrule
\end{tabular}
\vspace{-4pt}
\end{wraptable}

We choose baselines across three dense retrieval paradigms. BM25 is an unsupervised lexical baseline that serves as an exact-match ceiling. For dense bi-encoders, we selected four models differing in pre-training and architecture: (1) MiniLM-L6-v2 (speed-optimized, distilled 6-layer), (2) all-mpnet-base-v2 (mean-pooled MPNet with contrastive training on 1B+ pairs), (3) BGE-base-v1.5 (instruction-tuned bi-encoder from BAAI for zero-shot transfer), and (4) E5-base-v2 (weakly-supervised contrastive model with task-specific prompts). These span distilled vs. full-size and supervised vs. weakly supervised training, helping assess the impact of pre-training on CVE retrieval. Hybrid BM25+mpnet reranks the BM25 top-100 candidates with a weighted sum of normalized BM25 and dense cosine scores ($\alpha{=}0.7$ on the dense score), testing their complementarity. We measure Recall@$k$ ($ k= {10}$) and MRR, with results in Table~\ref{tab:task1-results}.

All dense models and the hybrid exceed BM25 (MRR 0.636) and cluster tightly (MRR 0.682--0.714). BGE-base-v1.5 has the highest MRR (0.714), but its difference from the hybrid (0.711) is not significant (paired $\Delta$MRR 0.003, 95\% CI [$-$0.013, 0.019]). No single model dominates: the hybrid leads at R@1 (0.679), and BGE and E5 lead at R@5 and R@10. Explicit identifiers are not the main signal: 24.8\% of test queries contain their gold CVE ID, yet removing CVE identifiers changes MRR by at most 0.004 for any model, and removing URLs, version strings, boilerplate, or all four changes it by at most 0.024 on either task (Appendix~\ref{app:shortcuts}). Verbatim overlap carries more of the signal: 918 of 1,990 test queries (46.1\%) share a span of at least 100 characters with their gold description. On the other 1,072, E5 (0.548) and BGE (0.541) lead, and the hybrid (0.480) and BM25 (0.373) fall behind (Table~\ref{tab:overlap}). Retrieval quality depends strongly on the query source. Hybrid MRR ranges from 0.95 on GitHub Advisory queries (1,036 of 1,990) to 0.29 on ExploitDB (429), where dense models lead BM25 (0.46 for BGE vs.\ 0.13).

\subsection{Task 2: Exploit Type Classification (ETC)}
\label{sec:task2}

Given an exploit text $x$, predict its vulnerability type $y \in \mathcal{Y}$ where $|\mathcal{Y}| = 8$ classes: \emph{injection} (SQLi, command/code injection, XXE), \emph{XSS} (reflected, stored, DOM), \emph{memory corruption} (buffer/heap overflow, use-after-free, format string), \emph{denial of service}, \emph{file inclusion} (LFI/RFI, path traversal, arbitrary file operations), \emph{authentication \& access control} (privilege escalation, auth bypass, CSRF, IDOR), \emph{remote code execution} (RCE, deserialization), and \emph{information disclosure} (SSRF, open redirect, data exposure). This taxonomy aligns with CWE top-level categories and prior exploit classification work~\citep{ampel2020dtl,ampel2024misq}. Labels are derived from structured metadata: ExploitDB exploit descriptions and tags, 0day.today exploit type fields, and HackerOne vulnerability type prefixes. The dataset comprises 70,883 labeled exploit texts from these three source families.

We use a temporal out-of-distribution split (train $<$2020, val 2020-2022, test $\geq$2023) so that models must generalize to vulnerabilities and exploits that postdate training. We evaluate a model ladder spanning bag-of-words classifiers (Decision Tree, TF-IDF+LR, linear SVM), recurrent neural networks (RNN, GRU, LSTM, BiLSTM), and three security-domain transformers (SecBERT~\citep{secbert2022}, SecureBERT~\citep{aghaei2023securebert}, CySecBERT~\citep{bayer2024cysecbert}). Every Task~2 model is trained with three seeds (42--44). We report macro-F1 and per-class F1 for three of the eight classes in Table~\ref{tab:task2-results}.

\begin{table}[t]
\centering
\small
\setlength{\tabcolsep}{4pt}
\caption{Task 2 (ETC) test results on 1,735 temporally held-out samples. Macro-F1 (mean $\pm$ SD, three seeds) and mean per-class F1 for three of eight classes. Bold = highest mean per column. BoW = bag-of-words; Transf. = security-domain transformers fine-tuned on a 15K subsample.}
\label{tab:task2-results}
\begin{tabular}{llcccc}
\toprule
& \textbf{Model} & \textbf{Macro-F1} & \textbf{info\_disc.} & \textbf{mem\_corr.} & \textbf{auth.} \\
\midrule
\multirow{3}{*}{\rotatebox[origin=c]{90}{\footnotesize BoW}}
& Decision Tree     & .702{\scriptsize$\pm$.007} & .45 & .64 & .60 \\
& TF-IDF + LR       & .772{\scriptsize$\pm$.000} & .66 & .69 & .79 \\
& SVM (Linear)      & .799{\scriptsize$\pm$.000} & .70 & .68 & .82 \\
\midrule
\multirow{4}{*}{\rotatebox[origin=c]{90}{\footnotesize RNNs}}
& RNN               & .436{\scriptsize$\pm$.031} & .30 & .42 & .44 \\
& GRU               & .858{\scriptsize$\pm$.001} & .70 & .84 & .81 \\
& LSTM              & .824{\scriptsize$\pm$.013} & .63 & .77 & .78 \\
& BiLSTM            & .856{\scriptsize$\pm$.003} & .69 & .84 & .81 \\
\midrule
\multirow{3}{*}{\rotatebox[origin=c]{90}{\footnotesize Transf.}}
& SecBERT           & .852{\scriptsize$\pm$.007} & .67 & .82 & .83 \\
& SecureBERT        & .892{\scriptsize$\pm$.008} & \textbf{.80} & .90 & .85 \\
& CySecBERT         & \textbf{.898}{\scriptsize$\pm$.003} & .79 & \textbf{.90} & \textbf{.86} \\
\bottomrule
\end{tabular}
\end{table}

The best bag-of-words baseline (SVM) attains macro-F1 of 0.799. Gated recurrent networks improve on it: GRU (0.858) and BiLSTM (0.856) are statistically indistinguishable across seeds, and LSTM reaches 0.824. The vanilla RNN performs poorly (macro-F1=0.436), indicating that long exploit texts require gated architectures. Security-domain pre-training helps most: SecureBERT (0.892) and CySecBERT (0.898) are the strongest models, and their difference is not significant (paired $\Delta$ 0.006, 95\% CI [$-$0.003, 0.016]), while SecBERT (0.852) is on par with the gated recurrent networks. Domain-adapted transformers therefore set the Task~2 baseline, even when fine-tuned on a 15K subsample. Information disclosure is the hardest class for every model (F1 at most 0.80). Disclosure exploits often involve injection (SSRF, XXE), while privilege escalation and authentication share patterns with RCE. Decision Tree drops to macro-F1=0.702, with disclosure at 0.45. Lexical classes such as XSS are easy for all models except the vanilla RNN (F1 $\geq$ 0.91), indicating that exploit signals are partly lexical. Temporal splits make the task harder as exploit names and vulnerability patterns evolve, preventing models from memorizing era-specific features.

\subsection{Task 3: Temporal Generalization (TG)}
\label{sec:task3}

\begin{wraptable}{r}{0.52\textwidth}
\vspace{-4pt}
\centering
\small
\setlength{\tabcolsep}{3pt}
\caption{Task 3 (TG) results (1,898 test queries) under strict CVE disjointness
($C_{\text{train}} \cap C_{\text{test}} = \emptyset$). Bootstrap 95\% MRR@10 intervals have a half-width of about 0.02; the top two MRR values do not differ significantly.}
\label{tab:task3-results}
\begin{tabular}{lcccc}
\toprule
\textbf{Model} & \textbf{R@1} & \textbf{R@5} & \textbf{R@10} & \textbf{MRR} \\
\midrule
BM25 & 0.620 & 0.713 & 0.740 & 0.660 \\
MiniLM-L6-v2 & 0.670 & 0.788 & 0.820 & 0.722 \\
mpnet-base-v2 & 0.647 & 0.777 & 0.809 & 0.703 \\
BGE-base-v1.5 & 0.667 & 0.798 & 0.828 & 0.726 \\
E5-base-v2 & 0.663 & \textbf{0.799} & \textbf{0.830} & 0.721 \\
\midrule
Hybrid BM25+mpnet & \textbf{0.698} & 0.772 & 0.786 & \textbf{0.730} \\
\bottomrule
\end{tabular}
\vspace{-4pt}
\end{wraptable}

Identical retrieval formulation to Task~1, but with a strict CVE-disjointness constraint: $C_{\text{train}} \cap C_{\text{test}} = \emptyset$, where $C_s = \{c^*_i : q_i \in \text{split}~s\}$ is the set of gold CVE targets in split $s$. Splits are defined by CVE publication year rather than query timestamp: train uses CVEs published before 2022, validation uses CVEs published in 2022--2023, and test uses CVEs published in 2024+. This ensures that models cannot memorize vulnerability descriptions seen during training and must generalize to wholly unseen vulnerabilities. We use the same metrics and models as Task~1 and show the results of Task~3 in Table~\ref{tab:task3-results}.

Under the CVE-disjoint evaluation, the hybrid has the highest MRR (0.730), but its difference from BGE-base-v1.5 (0.726) is not significant (paired $\Delta$MRR 0.004, 95\% CI [$-$0.012, 0.020]). Every model scores slightly higher than on Task~1 (e.g., BM25 MRR 0.660 vs.\ 0.636). The Task~3 test queries are a subset of the Task~1 test queries, so the two tasks differ in the constraint they place on training rather than in the difficulty of the evaluation. Fine-tuning BGE-base-v1.5 on the Task~3 training pairs raises test MRR to 0.757 ($\Delta$ 0.031, 95\% CI [0.019, 0.044]; Table~\ref{tab:b6}).

\begin{figure}[t]
\centering
\includegraphics[width=0.88\linewidth]{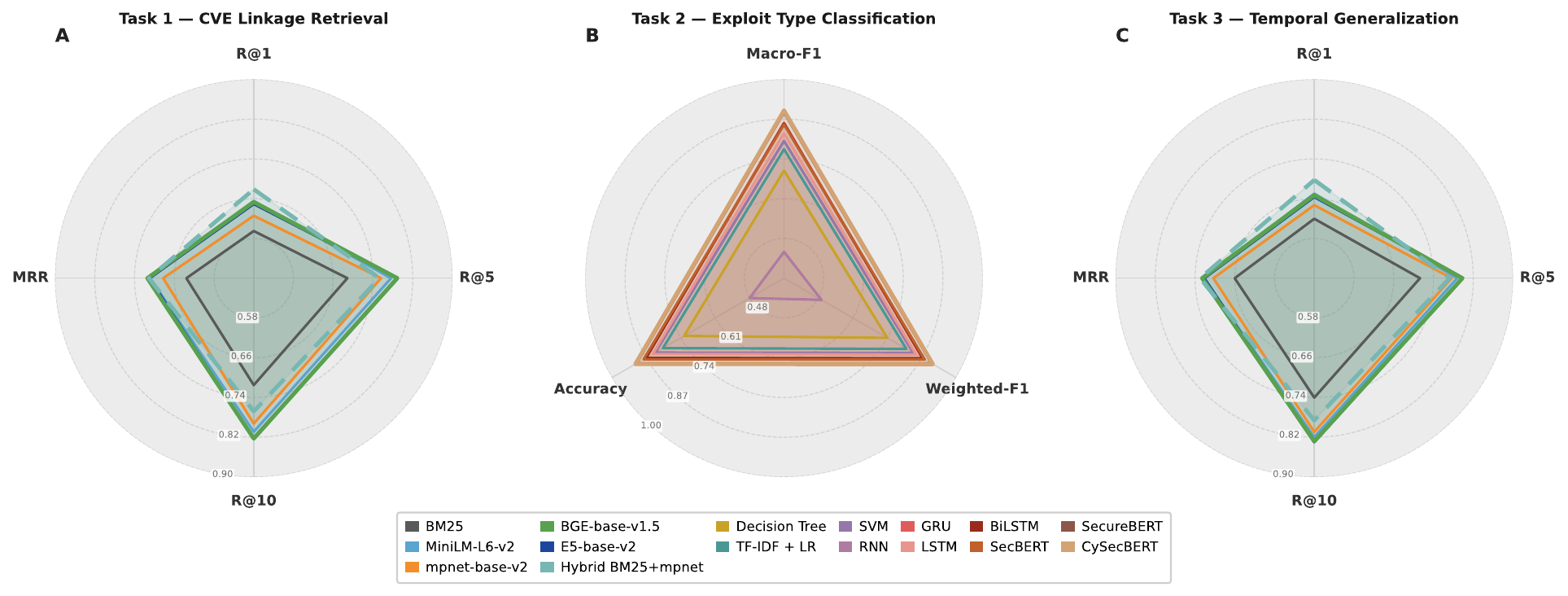}
\caption{Radar chart summary of baseline performance across all three tasks.}
\label{fig:baselines-summary}
\end{figure}

\subsection{Audit Artifacts and Error Analysis}

Following~\citet{kapoor2023leakage}, we verify Task~3's CVE-disjointness programmatically ($|C_{\text{train}} \cap C_{\text{test}}| = 0$; Appendix~\ref{app:audit-reliability}). Among the Hybrid model's Task~1 errors (queries where the target CVE does not appear in the top-10), we identify three recurring failure patterns: (1)~\emph{Generic exploit descriptions} matching multiple CVEs sharing a common vulnerability class; (2)~\emph{Version-specific queries} referencing product versions absent from the NVD description; and (3)~\emph{Alias mismatches} where queries use informal names (e.g., Log4Shell) while the NVD entry uses only formal identifiers.

\subsection{Forum-Post Grounding Pilot}
\label{sec:pilot}

The three tasks draw queries from structured sources. To test the discourse layer directly, we take the 1,964 hacker-community posts that cite a CVE, mask every CVE string, and ask each zero-shot retriever to recover the cited CVE from the same 340,536 NVD descriptions (Table~\ref{tab:pilot} in Appendix~\ref{app:shortcuts}). The best model, BGE, reaches MRR 0.078 (95\% CI [0.068, 0.089]), about nine times lower than on Task~1 (hybrid 0.711), and the five models differ by at most 0.014. Forum posts rarely restate the advisory text, and 30.2\% of them are Cyrillic-heavy. Grounding informal discourse to vulnerabilities is therefore far from solved.

\section{Ethical Considerations}
\label{sec:ethics}

The source mix contains various licenses and terms of service. The public release categorizes sources into redistributable, research-only, metadata/pointer, and excluded. A governance addendum details release modes, contact info, versioning, checksums, and takedown procedures. All sources are accessible without authentication, but privacy risks remain. We pseudonymize author IDs with a forum-scoped keyed HMAC-SHA-256 under an unpublished key, preserve provenance, and support takedown requests, downgrading ambiguous sources accordingly.
Where a source's terms in Table~\ref{tab:collection-details} conflict with the repository-level license tag, the per-source terms govern. HackerSignal highlights vulnerabilities for defensive research only, prohibiting fine-tuning automated exploit models, de-anonymizing forum users, and automated incident response without human review. Researchers should consult IRBs when analyzing individuals. Outputs should not be trusted as the current CTI without validation.

\section{Conclusion}
\label{sec:discussion}

HackerSignal links 7.45M documents from 64 sources across the CVE lifecycle with temporal splits, CVE-disjoint evaluation, and provenance-aware governance. Our three-task benchmark and model ladder (BM25 through dense and hybrid retrieval at MRR~0.71, security-domain transformers at macro-F1~0.90 for 8-class exploit type classification) establishes baselines for cross-source CVE grounding, exploit type classification, and prospective generalization to unseen vulnerabilities.

We note three limitations. First, CVE linkages are metadata-derived (source URLs, CVE regex extraction) rather than expert-adjudicated. The two-annotator audit finds 79.5\% linkage precision overall, ranging from 100\% (CVEfixes, GitHub Advisory) to 20.0\% (ZeroScience), so we report per-source reliability (Table~\ref{tab:source-reliability}). Second, the 6.97M-row hacker-community layer yields only about 2,000 CVE-linked records (0.03\% of the layer). Pre-disclosure discussion that never names a CVE cannot be evaluated without ground truth linking a post to a later disclosure. Implicit discourse-to-CVE grounding is therefore the main research direction the release enables, and the masked forum-post pilot (Section~\ref{sec:pilot}) gives it a first baseline. Third, 30.2\% of CVE-linked community records are Cyrillic-heavy, so discourse grounding in this corpus is partly a cross-lingual problem, and we report no cross-lingual baselines~\citep{ebrahimi2020crosslingual}. We also leave large language model rerankers and retrieval-augmented generation, which may score higher at far greater cost, to future work.

\section*{Acknowledgments}
This material is based upon work supported by the U.S. Army Research Laboratory (ARL) under Cooperative Agreement Number W911NF-23-2-0224. The views and conclusions contained in this document are those of the authors and should not be interpreted as representing the official policies, either expressed or implied, of the Army Research Laboratory or the U.S. Government. We acknowledge the use of the Advanced Research Computing Technology Innovation and Collaboration (ARCTIC) resources at Georgia State University. These resources were supported by the National Science Foundation through the Major Research Instrumentation (MRI) grant CNS-1920024 and the Research Infrastructure: CC Compute-Campus award for Pioneering Research-Oriented Flexible Cyberinfrastructure (PROFLEX-CI), grant 2430193. Sagar Samtani was supported in part by the National Science Foundation under grants CPIS-2338479 (CAREER) and 2613347 (CICI). The authors declare no competing interests.

\appendix

\section{Source Collection Details}
\label{app:collection}

The full source-by-source matrix and takedown/correction process are in \texttt{docs/release\_governance.md}. Collection scripts are available at \href{https://github.com/BenAmpel/hackersignal}{\texttt{github.com/BenAmpel/hackersignal}} under \texttt{src/etg/collectors/}. The public JSONL release is built from the full research build by enforcing \texttt{release\_mode}: rows from \texttt{metadata\_or\_pointer\_only} sources retain provenance, timestamps, source/thread identifiers, CVE references, text hashes, and text lengths, but \texttt{text} and \texttt{text\_raw} are withheld. Table~\ref{tab:collection-details} provides per-source collection method, access mechanism, license posture, and default release mode.

\begin{table}[t]
\centering
\small
\caption{Per-source collection details.}
\label{tab:collection-details}
\begin{tabular}{lllll}
\toprule
\textbf{Source} & \textbf{Method} & \textbf{Access} & \textbf{License} & \textbf{Release mode} \\
\midrule
NVD             & REST API        & Public & Public domain & Text \\
CISA KEV        & JSON feed       & Public & Public domain & Text \\
GitHub Advisory & GraphQL API     & Public & CC BY 4.0 & Text \\
ExploitDB       & CSV download    & Public & CC BY-SA 4.0 & Text \\
HackerOne       & HuggingFace Hub & Public & Source dataset terms & Research text \\
CVEfixes        & Zenodo SQLite   & Public & CC BY 4.0 & Text \\
Kaggle forums   & Kaggle          & Public & Dataset-specific & Research/pointer \\
DeepDarkCTI     & GitHub          & Public & Repository terms & Research text \\
0x00sec/HTB/etc.& Discourse API   & Public & Site terms & Pointer \\
Other web archives & Web scraping & Public & Mixed terms & Pointer \\
\bottomrule
\end{tabular}
\end{table}

\section{Audit and Reliability}
\label{app:audit-reliability}

We follow the audit framework of~\citet{kapoor2023leakage} and perform three complementary integrity checks. First, for each task, we draw an 80-row stratified random sample (240 rows total, balanced across splits) and provide a structured annotation packet in \texttt{data/audits/} covering (i)~correct CVE assignment, (ii)~temporal split adherence, and (iii)~absence of trivial textual overlap between query and target. Programmatic checks on the sampling frame found no structural errors. Second, we search for near-duplicate text between the training and test splits of each task with MinHash over character 3-shingles (128 permutations, LSH at Jaccard $\geq$0.8, every candidate confirmed by exact Jaccard). No Task~2 test row has a training match. Two Task~1 test queries (of 1,990) have one, at Jaccard 0.80 and 0.82; both are also Task~3 test queries, where their training matches describe different CVEs. Finally, we check gold-CVE overlap between splits. For Task~3 (temporal generalization), CVE-disjointness is enforced by construction: $|C_{\text{train}} \cap C_{\text{test}}| = 0$, supported programmatically at build time. Task~1 does not impose disjointness; 9 of its 1,876 test CVEs (9 of 1,990 test queries) also occur in training.

\section{Model Ladder Details}
\label{app:model-ladder}

The full model ladder evaluates six retrieval approaches (Tasks~1 and~3) and ten classification approaches (Task~2). Each model represents a rung on the complexity ladder from lexical to neural to hybrid methods. BM25 uses no learned parameters; bi-encoders (MiniLM, mpnet, BGE, E5) use pre-trained sentence embeddings without task-specific fine-tuning; the hybrid combines BM25 top-100 reranking with dense scores ($\alpha{=}0.7$). For Task~2, the ladder spans bag-of-words classifiers (Decision Tree, TF-IDF + LR, linear SVM), recurrent neural networks (RNN, GRU, LSTM, BiLSTM), and three security-domain transformers (SecBERT, SecureBERT, CySecBERT).

\section{Baseline Hyperparameter Details}
\label{app:hyperparams}

\paragraph{BM25 (Tasks 1 and 3).} 
Okapi BM25 via \texttt{rank\_bm25} (BM25Okapi). Tokenization: lowercase whitespace split. Full 340,536-record corpus indexed.

\paragraph{Dense bi-encoders (Tasks 1 and 3).}
Four sentence-transformer models evaluated without task-specific fine-tuning: \texttt{all-MiniLM-L6-v2} (22M params, 384-dim), \texttt{all-mpnet-base-v2} (110M, 768-dim), \texttt{BAAI/bge-base-en-v1.5} (110M, 768-dim), and \texttt{intfloat/e5-base-v2} (110M, 768-dim, with ``query:''/``passage:'' prefixes). Inputs are the first 384 characters of each query and document, max sequence length 512 tokens, batch size 128, $\ell_2$-normalized fp32 embeddings, cosine similarity scoring; BGE is used without its query instruction. MRR is computed over the top 10 (MRR@10), with ties broken by corpus order. Two settings matter: encoding full texts instead of the first 384 characters raises Task~1 MRR for the base-size models by about 0.03 (e.g., BGE 0.744, E5 0.741) and lowers MiniLM by 0.025, and adding BGE's query instruction raises its MRR to 0.725 (Task~1) and 0.737 (Task~3). E5's prefixes change MRR by less than 0.001.

\paragraph{Hybrid BM25 + dense (Tasks 1 and 3).}
A combination of normalized BM25 scores and dense cosine similarity with $\alpha{=}0.7$ (dense weight). For efficiency, BM25 retrieves the top 100 candidates per query; dense scores are computed only for the candidate set. Final ranking uses $\alpha \cdot \text{dense} + (1{-}\alpha) \cdot \text{BM25}$.

\paragraph{Bag-of-words classifiers (Task 2).}
All BoW classifiers use scikit-learn with \texttt{TfidfVectorizer}(max\_features=50{,}000, ngram\_range=(1,2)) as the feature extractor. Features are computed from the first 2{,}000 characters of each text. \texttt{DecisionTreeClassifier}: max\_depth=30, class\_weight=``balanced''. \texttt{LogisticRegression}: C=1.0, max\_iter=1000, class\_weight=``balanced''. \texttt{LinearSVC}: max\_iter=2000, class\_weight=``balanced''.

\paragraph{Neural classifiers (Task 2).}
RNN, GRU, LSTM, and BiLSTM use a lower-cased, whitespace-tokenized word-level vocabulary (minimum frequency $\geq$3, yielding 175,062 tokens) with 128-dim embeddings, 128-dim hidden states, max sequence length 300 (right-padded), batch size 128, and Adam optimizer (lr=$10^{-3}$). Class-weighted cross-entropy loss with gradient clipping at 1.0. Sequences are packed so that the final hidden state is taken at the last non-padding token. Trained for 8 epochs with the best epoch selected by validation macro-F1. SecBERT~\citep{secbert2022} (\texttt{jackaduma/SecBERT}), SecureBERT~\citep{aghaei2023securebert} (\texttt{ehsanaghaei/SecureBERT}), and CySecBERT~\citep{bayer2024cysecbert} (\texttt{markusbayer/CySecBERT}) are fine-tuned with max sequence length 256 on the first 1{,}000 characters of each text, batch size 16, AdamW (lr=$2{\times}10^{-5}$, weight decay 0.01), linear warmup over 10\% of steps, and unweighted cross-entropy, for 3 epochs on a 15K training subsample, with the best epoch selected by validation macro-F1. Every Task~2 model is trained with seeds 42, 43, and 44 (the 15K subsample is drawn per seed); we report the mean and SD. DarkBERT~\citep{jin2023darkbert} was not run: the model is gated, and our access request was not granted before the camera-ready deadline. All neural models use MPS (Apple Silicon) acceleration.

\paragraph{Compute.}
All baselines ran on one Apple M5 Pro workstation (48\,GB unified memory, macOS 27) with Python 3.12.15, PyTorch 2.14.1 (MPS backend), sentence-transformers 6.2.0, transformers 5.21.0, and scikit-learn 1.9.1. The only exception is the full-training-set transformer runs (Table~\ref{tab:task2-full}), which used one NVIDIA L40S GPU on the Georgia State University ACIDS cluster (PyTorch 2.14.1, CUDA 12.6): about 12\,min (SecBERT) to 22\,min (SecureBERT, CySecBERT) per seed. BM25 uses an exact sparse reimplementation of \texttt{rank\_bm25} (maximum score difference below $3{\times}10^{-12}$ on a 50-query check): 5--8\,s to index the 340K corpus and 15--25\,s to score each test set. Dense encoding of the 340K corpus took ${\sim}$4\,min (MiniLM) to ${\sim}$30\,min (base-sized models). Task~2 training per seed: BoW under 30\,s; RNN ${\sim}$30\,min, GRU ${\sim}$36\,min, LSTM ${\sim}$14\,min, BiLSTM ${\sim}$61\,min; SecBERT ${\sim}$17\,min, SecureBERT ${\sim}$40\,min, and CySecBERT ${\sim}$44\,min (15K subsample, 3 epochs). Per-run timings, seeds, model revisions, and data hashes are in \texttt{results/camera\_ready/provenance.json}. All baseline code is available at \href{https://github.com/BenAmpel/hackersignal}{\texttt{github.com/BenAmpel/hackersignal}}.

\paragraph{Full training set (Task 2 transformers).}
Table~\ref{tab:task2-full} repeats the three transformers with all 64,413 training rows (same code, data, and protocol; best epoch by validation macro-F1; seeds 42--44). More training data lifts all three to 0.899--0.903 and removes SecBERT's gap on the 15K subsample; the seed SDs overlap, so no transformer leads on the full training set.

\begin{table}[ht]
\centering
\small
\setlength{\tabcolsep}{4pt}
\caption{Task 2 transformers fine-tuned on the 15K subsample (Apple M5 Pro, Table~\ref{tab:task2-results}) and on all 64,413 training rows (one NVIDIA L40S). Macro-F1 mean $\pm$ SD over three seeds; per-class F1 for the full training set.}
\label{tab:task2-full}
\begin{tabular}{lccccc}
\toprule
\textbf{Model} & \textbf{15K subsample} & \textbf{Full 64K} & \textbf{info\_disc.} & \textbf{mem\_corr.} & \textbf{auth.} \\
\midrule
SecBERT & 0.852{\scriptsize$\pm$0.007} & 0.899{\scriptsize$\pm$0.001} & 0.75 & 0.90 & 0.88 \\
SecureBERT & 0.892{\scriptsize$\pm$0.008} & 0.903{\scriptsize$\pm$0.006} & 0.81 & 0.90 & 0.88 \\
CySecBERT & 0.898{\scriptsize$\pm$0.003} & 0.901{\scriptsize$\pm$0.011} & 0.81 & 0.88 & 0.88 \\
\bottomrule
\end{tabular}
\end{table}

\paragraph{Fine-tuned bi-encoders (Task 3).}
We fine-tune BGE-base-v1.5 and MiniLM-L6-v2 on the Task~3 training split: 56,833 (query, gold NVD description) pairs whose CVEs were published before 2022, both sides truncated to 384 characters, dropping pairs whose two texts are identical. Training runs one epoch of multiple-negatives ranking loss with in-batch negatives (batch 64 computed in mini-batches of 16, scale 20, no repeated CVE within a batch), learning rate $2\times10^{-5}$, 10\% warmup, fp32, seeds 42--44, on the Apple M5 Pro. No model selection is done: the validation split (2022--2023 CVEs) is reported in the released results but not used. Evaluation follows the zero-shot protocol above over all 340,536 descriptions (Table~\ref{tab:b6}).

\begin{table}[ht]
\centering
\small
\setlength{\tabcolsep}{4pt}
\caption{Task 3 test MRR@10 for bi-encoders zero-shot and fine-tuned on the Task~3 training split. Fine-tuned: mean $\pm$ SD over three seeds, with a bootstrap 95\% CI of the seed-mean MRR; $\Delta$ is the paired bootstrap difference from zero-shot (10,000 resamples).}
\label{tab:b6}
\begin{tabular}{lccccc}
\toprule
\textbf{Model} & \textbf{Zero-shot} & \textbf{Fine-tuned} & \textbf{95\% CI} & \textbf{$\Delta$MRR [95\% CI]} & \textbf{R@1} \\
\midrule
BGE-base-v1.5 & 0.726 & 0.757{\scriptsize$\pm$0.005} & [0.740, 0.774] & 0.031 [0.019, 0.044] & 0.667 $\to$ 0.693 \\
MiniLM-L6-v2 & 0.722 & 0.750{\scriptsize$\pm$0.002} & [0.732, 0.767] & 0.028 [0.016, 0.039] & 0.670 $\to$ 0.688 \\
\bottomrule
\end{tabular}
\end{table}

\section{Shortcut and Source Diagnostics}
\label{app:shortcuts}

Table~\ref{tab:pilot} gives the forum-post grounding pilot (Section~\ref{sec:pilot}) for every zero-shot retriever. Table~\ref{tab:masking} reports MRR for every retrieval model after removing one shortcut family from the queries: CVE identifiers, URLs, version strings, or boilerplate, and all four together. In Task~1, 24.8\% of test queries contain their gold CVE identifier and 49.1\% contain a URL. Table~\ref{tab:per-source} stratifies Task~1 MRR by query source for every model. Task~3 shares these sources and gives nearly identical per-source values (released in \texttt{results/camera\_ready/results.csv}). Table~\ref{tab:overlap} splits each task's test queries by verbatim overlap with the gold description (a shared span of at least 100 characters; 918 of 1,990 Task~1 and 911 of 1,898 Task~3 queries). Removing from each query every span of 30 or more characters it shares with its gold description lowers BM25 MRR from 0.636 to 0.321 on Task~1 and from 0.660 to 0.335 on Task~3. This stripping uses the gold description, so it is a diagnostic bound rather than a deployable condition.

\begin{table}[ht]
\centering
\small
\setlength{\tabcolsep}{4pt}
\caption{Test MRR@10 by verbatim overlap with the gold NVD description (a shared span of $\geq$100 characters). $^\dagger$BM25 with every span of $\geq$30 characters shared with the gold description removed from the query (a diagnostic; it uses the gold description).}
\label{tab:overlap}
\begin{tabular}{lcccc}
\toprule
 & \multicolumn{2}{c}{\textbf{Task 1}} & \multicolumn{2}{c}{\textbf{Task 3}} \\
\cmidrule(lr){2-3}\cmidrule(lr){4-5}
\textbf{Model} & \textbf{Overlap} ($n$=918) & \textbf{None} ($n$=1,072) & \textbf{Overlap} ($n$=911) & \textbf{None} ($n$=987) \\
\midrule
BM25 & 0.942 & 0.373 & 0.943 & 0.399 \\
MiniLM & 0.925 & 0.523 & 0.926 & 0.534 \\
mpnet & 0.899 & 0.497 & 0.903 & 0.519 \\
BGE & 0.916 & 0.541 & 0.918 & 0.549 \\
E5 & 0.899 & 0.548 & 0.900 & 0.556 \\
Hybrid & 0.981 & 0.480 & 0.982 & 0.497 \\
\midrule
BM25, overlap stripped$^\dagger$ & \multicolumn{2}{c}{0.321} & \multicolumn{2}{c}{0.335} \\
\bottomrule
\end{tabular}
\end{table}

\begin{table}[ht]
\centering
\small
\setlength{\tabcolsep}{3pt}
\caption{Masked forum-post grounding pilot (1,964 posts, CVE strings masked, 340K corpus).}
\label{tab:pilot}
\begin{tabular}{lccc}
\toprule
\textbf{Model} & \textbf{R@1} & \textbf{R@10} & \textbf{MRR} \\
\midrule
BM25 & 0.046 & 0.109 & 0.064 \\
MiniLM & 0.053 & 0.122 & 0.073 \\
mpnet & 0.041 & 0.127 & 0.065 \\
BGE & 0.056 & 0.132 & \textbf{0.078} \\
E5 & 0.049 & 0.125 & 0.072 \\
\bottomrule
\end{tabular}
\end{table}

\begin{table}[ht]
\centering
\small
\setlength{\tabcolsep}{4pt}
\caption{Shortcut masking: test MRR@10 with each shortcut family removed from the queries.}
\label{tab:masking}
\begin{tabular}{lcccccc}
\toprule
\textbf{Model} & \textbf{As rel.} & \textbf{$-$CVE} & \textbf{$-$URL} & \textbf{$-$Ver.} & \textbf{$-$Boil.} & \textbf{All} \\
\midrule
\multicolumn{7}{l}{\textit{Task 1}} \\
BM25 & 0.636 & 0.635 & 0.635 & 0.630 & 0.648 & 0.642 \\
MiniLM & 0.709 & 0.712 & 0.710 & 0.702 & 0.726 & 0.719 \\
mpnet & 0.682 & 0.686 & 0.686 & 0.669 & 0.707 & 0.694 \\
BGE & 0.714 & 0.717 & 0.718 & 0.698 & 0.724 & 0.714 \\
E5 & 0.710 & 0.711 & 0.711 & 0.691 & 0.729 & 0.703 \\
Hybrid & 0.711 & 0.712 & 0.711 & 0.702 & 0.710 & 0.705 \\
\midrule
\multicolumn{7}{l}{\textit{Task 3}} \\
BM25 & 0.660 & 0.660 & 0.659 & 0.655 & 0.671 & 0.667 \\
MiniLM & 0.722 & 0.726 & 0.724 & 0.716 & 0.736 & 0.732 \\
mpnet & 0.703 & 0.705 & 0.704 & 0.690 & 0.719 & 0.707 \\
BGE & 0.726 & 0.727 & 0.727 & 0.713 & 0.727 & 0.722 \\
E5 & 0.721 & 0.722 & 0.721 & 0.704 & 0.735 & 0.714 \\
Hybrid & 0.730 & 0.730 & 0.730 & 0.723 & 0.730 & 0.728 \\
\bottomrule
\end{tabular}
\end{table}

\begin{table}[ht]
\centering
\small
\setlength{\tabcolsep}{4pt}
\caption{Task~1 test MRR@10 by query source.}
\label{tab:per-source}
\begin{tabular}{lrcccccc}
\toprule
\textbf{Source} & \textbf{$n$} & \textbf{BM25} & \textbf{MiniLM} & \textbf{mpnet} & \textbf{BGE} & \textbf{E5} & \textbf{Hybrid} \\
\midrule
GitHub Advisory & 1036 & 0.901 & 0.926 & 0.905 & 0.917 & 0.888 & 0.950 \\
ExploitDB & 429 & 0.134 & 0.435 & 0.363 & 0.463 & 0.463 & 0.289 \\
CISA KEV & 363 & 0.555 & 0.558 & 0.548 & 0.564 & 0.611 & 0.615 \\
ZeroScience & 102 & 0.603 & 0.503 & 0.547 & 0.571 & 0.577 & 0.683 \\
Seebug & 38 & 0.089 & 0.043 & 0.053 & 0.022 & 0.037 & 0.097 \\
HackerOne & 17 & 0.382 & 0.543 & 0.433 & 0.522 & 0.637 & 0.549 \\
0x00sec & 5 & 0.300 & 0.029 & 0.000 & 0.000 & 0.050 & 0.167 \\
\bottomrule
\end{tabular}
\end{table}

\section{Comparison with Prior CTI Datasets}
\label{app:comparison}

Table~\ref{tab:dataset-comparison} positions HackerSignal against prior cyber threat intelligence datasets. We emphasize four differentiators: (i) governed public release with source-specific access modes; (ii) temporal benchmark splits enabling prospective evaluation; (iii) cross-source CVE linkage connecting hacker discourse to formal vulnerability records; and (iv) the largest public CVE-linked, multi-source CTI release. CrimeBB is larger in raw posts (48M) but is credentialed rather than openly released and carries no CVE linkage, temporal splits, or benchmark tasks~\citep{crimebb2018}.

\begin{table}[t]
\centering
\small
\setlength{\tabcolsep}{3pt}
\caption{Comparison of HackerSignal with prior CTI datasets across key
dimensions. \cmark{} = yes, \xmark{} = no, $\sim$ = partial.}
\label{tab:dataset-comparison}
\resizebox{\linewidth}{!}{%
\begin{tabular}{lccccccc}
\toprule
\textbf{Dataset} & \textbf{Size} & \textbf{Sources} & \textbf{Public} & \textbf{CVE Link} & \textbf{Temp.\ Splits} & \textbf{Governance} & \textbf{Benchmark} \\
\midrule
CrimeBB~\citep{crimebb2018} & 48M posts & 4 forums & \xmark{} & \xmark{} & \xmark{} & \xmark{} & \xmark{} \\
Gayanku~\citep{gayanku2020} & 3.8M posts & 7 markets & \cmark{} & \xmark{} & \xmark{} & \xmark{} & \xmark{} \\
Otto et al.~\citep{otto2021isi} & 259K posts & 3 forums & \cmark{} & \xmark{} & \xmark{} & \xmark{} & \xmark{} \\
ExploitDB~\citep{exploitdb} & 50K+ & 1 archive & \cmark{} & $\sim$ & \xmark{} & \cmark{} & \xmark{} \\
CVEfixes~\citep{cvefixes2021} & 8.3K & commits & \cmark{} & \cmark{} & \xmark{} & \cmark{} & \xmark{} \\
\midrule
\textbf{HackerSignal (ours)} & \textbf{7.45M} & \textbf{64 sources} & \cmark{} & \cmark{} & \cmark{} & \cmark{} & \cmark{} \\
\bottomrule
\end{tabular}
}
\end{table}

\section{Datasheet}
\label{app:datasheet}

A complete datasheet is provided in the supplemental material and in the dataset repository at \texttt{docs/datasheet.md}~\citep{gebru2021datasheet}.

\newpage
\section*{NeurIPS Paper Checklist}

\begin{enumerate}

\item {\bf Claims}
    \item[] Question: Do the main claims made in the abstract and introduction accurately reflect the paper's contributions and scope?
    \item[] Answer: \answerYes{}
    \item[] Justification: The abstract states three concrete contributions (corpus, benchmark tasks, baselines) that are fully realized in Sections~\ref{sec:dataset} and~\ref{sec:tasks}.

\item {\bf Limitations}
    \item[] Question: Does the paper discuss the limitations of the work performed by the authors?
    \item[] Answer: \answerYes{}
\item[] Justification: Section~\ref{sec:dataset} discusses data quality (duplicates, temporal imbalance, label noise) and Section~\ref{sec:discussion} states three limitations (metadata-derived labels with per-source reliability, sparse CVE-linked discourse, and the cross-lingual share of that discourse) and the missing LLM baselines.

\item {\bf Theory assumptions and proofs}
    \item[] Question: For each theoretical result, does the paper provide the full set of assumptions and a complete proof?
    \item[] Answer: \answerNA{}
    \item[] Justification: This is a dataset paper with no theoretical results.

\item {\bf Experimental result reproducibility}
    \item[] Question: Does the paper fully disclose all information needed to reproduce the main experimental results?
    \item[] Answer: \answerYes{}
    \item[] Justification: All baseline hyperparameters are reported in Appendix~\ref{app:hyperparams}. Code is released at \href{https://github.com/BenAmpel/hackersignal}{\texttt{github.com/BenAmpel/hackersignal}}. Data is publicly available on HuggingFace Hub.

\item {\bf Open access to data and code}
    \item[] Question: Does the paper provide open access to the data and code?
    \item[] Answer: \answerYes{}
    \item[] Justification: Dataset released on HuggingFace with provenance metadata, Croissant metadata, and a release-governance addendum. Collection and baseline code released at \href{https://github.com/BenAmpel/hackersignal}{\texttt{github.com/BenAmpel/hackersignal}}. Accessible without personal request to the PI, with pointer-only modes for sources whose terms do not support text mirroring.

\item {\bf Experimental setting/details}
    \item[] Question: Does the paper specify all training and test details necessary to understand the results?
    \item[] Answer: \answerYes{}
    \item[] Justification: Split boundaries, sources, and hyperparameters are specified in Section~\ref{sec:tasks} and Appendix~\ref{app:hyperparams}.

\item {\bf Experiment statistical significance}
    \item[] Question: Does the paper report error bars or statistical significance?
    \item[] Answer: \answerYes{}
    \item[] Justification: Tables~\ref{tab:task1-results} and~\ref{tab:task3-results} report test-set bootstrap 95\% intervals on MRR (10,000 resamples) and paired bootstrap tests for the top two models; Task~2 reports the mean $\pm$ SD of macro-F1 over three seeds (Table~\ref{tab:task2-results}). BM25 and dense-retrieval baselines are deterministic given a fixed corpus and queries (ties are broken by corpus order). MPS training is not bit-deterministic, which the seed SD captures.

\item {\bf Experiments compute resources}
    \item[] Question: Does the paper provide sufficient information on compute resources?
    \item[] Answer: \answerYes{}
    \item[] Justification: Baselines were run on one Apple M5 Pro workstation (48\,GB) with PyTorch 2.14.1 on the MPS backend; the full-training-set transformer runs used one NVIDIA L40S GPU. Appendix~\ref{app:hyperparams} reports wall-clock times per model and task.

\item {\bf Code of ethics}
    \item[] Question: Does the research conform with the NeurIPS Code of Ethics?
    \item[] Answer: \answerYes{}
    \item[] Justification: Data are public OSINT, but the paper treats public availability as insufficient by itself: authors are pseudonymized, ambiguous sources can be released pointer-only, and dual-use/takedown restrictions are stated in Section~\ref{sec:ethics} and the governance addendum.

\item {\bf Broader impacts}
    \item[] Question: Does the paper discuss both positive and negative societal impacts?
    \item[] Answer: \answerYes{}
    \item[] Justification: Section~\ref{sec:ethics} addresses dual-use risk, pseudonymization, and prohibited uses.

\item {\bf Safeguards}
    \item[] Question: Does the paper describe safeguards for responsible release?
    \item[] Answer: \answerYes{}
    \item[] Justification: Explicit prohibited uses listed in Section~\ref{sec:ethics} and the governance addendum. Author pseudonyms are keyed HMAC-SHA-256 digests applied at release time; disputed or terms-ambiguous sources can be downgraded to metadata/pointer-only release.

\item {\bf Licenses for existing assets}
    \item[] Question: Are existing assets properly credited with licenses?
    \item[] Answer: \answerYes{}
    \item[] Justification: Appendix~\ref{app:collection} lists the license for every source. The References section cites all dataset papers.

\item {\bf New assets}
    \item[] Question: Are new assets well documented?
    \item[] Answer: \answerYes{}
    \item[] Justification: A Gebru et al.\ datasheet is provided (Appendix~\ref{app:datasheet} and \texttt{docs/datasheet.md}). Croissant machine-readable metadata is distributed with the HuggingFace repository.

\item {\bf Crowdsourcing and research with human subjects}
    \item[] Question: Does the paper include details for crowdsourcing or research with human subjects?
    \item[] Answer: \answerNA{}
    \item[] Justification: No crowdsourcing was conducted. The release uses public OSINT sources, pseudonymizes authors, supports takedown/correction requests, and instructs researchers making individual-level inferences to seek their own IRB review.

\item {\bf Institutional review board (IRB) approvals}
    \item[] Question: Does the paper describe IRB approvals for research with human subjects?
    \item[] Answer: \answerNA{}
    \item[] Justification: Collection of public OSINT data was not classified as human-subjects research under institutional guidelines; the paper still documents privacy safeguards, responsible-use restrictions, and release-mode controls.

\item {\bf Declaration of LLM usage}
    \item[] Question: Does the paper describe the usage of LLMs if it is an important, original, or non-standard component of the core methods in this research? Note that if the LLM is used only for writing, editing, or formatting purposes and does not impact the core methodology, scientific rigor, or originality of the research, declaration is not required.
    \item[] Answer: \answerNA{}
    \item[] Justification: LLMs are not an important, original, or non-standard component of the core methods; they were used only for writing and editing assistance.

\end{enumerate}


\begin{thebibliography}{99}
\small

\bibitem[Akhtar et~al.(2024)]{croissant2024}
Akhtar, M., et al. (2024). Croissant: A Metadata Format
for ML-Ready Datasets. \textit{NeurIPS 2024 D\&B Track}.

\bibitem[Ampel et~al.(2020)]{ampel2020dtl}
Ampel, B., Samtani, S., Zhu, H., Ullman, S., \& Chen, H. (2020).
Labeling Hacker Exploits for Proactive Cyber Threat Intelligence: A Deep
Transfer Learning Approach. \textit{IEEE Intelligence and Security
Informatics (ISI)}. \url{https://doi.org/10.1109/ISI49825.2020.9280548}.

\bibitem[Ampel et~al.(2024)]{ampel2024misq}
Ampel, B., Samtani, S., Zhu, H., \& Chen, H. (2024).
Creating Proactive Cyber Threat Intelligence with Hacker Exploit Labels:
A Deep Transfer Learning Approach. \textit{MIS Quarterly}, 48(1), 137--166.
\url{https://doi.org/10.25300/MISQ/2023/17316}.

\bibitem[Robertson and Jones(1994)]{bm25robertson1994}
Robertson, S., \& Jones, K.\ S. (1994).
Simple, proven approaches to text retrieval.
\textit{Technical report}, University of Cambridge.

\bibitem[Thakur et~al.(2021)]{thakur2021beir}
Thakur, N., Reimers, N., R{\"u}ckl{\'e}, A., Srivastava, A., \&
Gurevych, I. (2021). BEIR: A Heterogeneous Benchmark for Zero-shot
Evaluation of Information Retrieval Models. \textit{NeurIPS 2021 D\&B
Track}. arXiv:2104.08663.

\bibitem[CISA(2021--2026)]{cisakev}
CISA. (2021--2026). Known Exploited Vulnerabilities Catalog.
\url{https://www.cisa.gov/known-exploited-vulnerabilities-catalog}.

\bibitem[Pastrana et~al.(2018)]{crimebb2018}
Pastrana, S., Thomas, D.\ R., Hutchings, A., \&
Clayton, R. (2018). CrimeBB: Enabling Cybercrime Research on Underground
Forums at Scale. \textit{WWW 2018}.

\bibitem[Deliu et~al.(2017)]{deliu2017cti}
Deliu, I., Leichter, C., \& Franke, K. (2017).
Extracting Cyber Threat Intelligence from Hacker Forums: Support Vector
Machines versus Convolutional Neural Networks.
\textit{IEEE International Conference on Big Data}.
\url{https://doi.org/10.1109/BigData.2017.8258359}.

\bibitem[Bhandari et~al.(2021)]{cvefixes2021}
Bhandari, G., Naseer, A., \& Moonen, L. (2021).
CVEfixes: Automated Collection of Vulnerabilities and Their Fixes from
Open-Source Software. \textit{PROMISE 2021}.

\bibitem[Devlin et~al.(2019)]{devlin2019bert}
Devlin, J., Chang, M.-W., Lee, K., \& Toutanova, K.
(2019). BERT: Pre-training of Deep Bidirectional Transformers for
Language Understanding. \textit{NAACL 2019}.

\bibitem[Ebrahimi et~al.(2020)]{ebrahimi2020crosslingual}
Ebrahimi, M., Samtani, S., Chai, Y., \& Chen, H. (2020).
Detecting Cyber Threats in Non-English Hacker Forums: An Adversarial
Cross-Lingual Knowledge Transfer Approach.
\textit{IEEE Security and Privacy Workshops}, 20--26.
\url{https://doi.org/10.1109/SPW50608.2020.00021}.

\bibitem[Offensive Security(2003--2026)]{exploitdb}
Offensive Security. (2003--2026). Exploit Database.
\url{https://www.exploit-db.com/}. Licensed CC BY-SA 4.0.

\bibitem[Gayanku(2020)]{gayanku2020}
Gayanku. (2020). Hacker Forum Posts Dataset.
\textit{Kaggle}. \url{https://www.kaggle.com/gayanku}.

\bibitem[Gebru et~al.(2021)]{gebru2021datasheet}
Gebru, T., et al. (2021). Datasheets for Datasets.
\textit{Communications of the ACM}, 64(12), 86--92.

\bibitem[Cohen(1960)]{cohen1960kappa}
Cohen, J. (1960). A coefficient of agreement for nominal scales.
\textit{Educational and Psychological Measurement}, 20(1), 37--46.
\url{https://doi.org/10.1177/001316446002000104}.

\bibitem[GitHub(2017--2026)]{githubadvisory}
GitHub. (2017--2026). GitHub Advisory Database.
\url{https://github.com/advisories}. Licensed CC BY 4.0.

\bibitem[Grisham et~al.(2017)]{grisham2017mobile}
Grisham, J., Samtani, S., Patton, M., \& Chen, H. (2017).
Identifying Mobile Malware and Key Threat Actors in Online Hacker Forums
for Proactive Cyber Threat Intelligence.
\textit{IEEE Intelligence and Security Informatics (ISI)}.

\bibitem[Otto et~al.(2021)]{otto2021isi}
Otto, K., Ampel, B., Zhu, H., Samtani, S., \& Chen, H. (2021).
Exploring the Evolution of Exploit-Sharing Hackers: An Unsupervised
Graph Embedding Approach. \textit{IEEE Intelligence and Security
Informatics (ISI)}. \url{https://doi.org/10.1109/ISI53945.2021.9624846}.

\bibitem[NIST(2002--2026)]{nvd}
NIST. (2002--2026). National Vulnerability Database.
\url{https://nvd.nist.gov/}. Public domain.

\bibitem[Petroni et~al.(2021)]{petroni2021kilt}
Petroni, F., Piktus, A., Fan, A., Lewis, P., Yazdani, M., et al.
(2021). KILT: A Benchmark for Knowledge Intensive Language Tasks.
\textit{NAACL 2021}. \url{https://doi.org/10.18653/v1/2021.naacl-main.200}.

\bibitem[Nunes et~al.(2016)]{nunes2016darknet}
Nunes, E., Diab, A., Gunn, A., Marin, E., Mishra, V.,
Paliath, V., Robertson, J., Shakarian, J., Thart, A., \&
Shakarian, P. (2016). Darknet and Deepnet Mining for Proactive
Cybersecurity Threat Intelligence. \textit{IEEE Intelligence and
Security Informatics (ISI)}.

\bibitem[Jackaduma(2022)]{secbert2022}
Jackaduma. (2022). SecBERT: A Pre-trained Language Model
for Cybersecurity. \url{https://huggingface.co/jackaduma/SecBERT}.

\bibitem[Aghaei et~al.(2023)]{aghaei2023securebert}
Aghaei, E., Niu, X., Shadid, W., \& Al-Shaer, E. (2023). SecureBERT: A Domain-Specific Language Model for Cybersecurity. In \emph{Security and Privacy in Communication Networks (SecureComm 2022)}, pp. 39--56. Springer.

\bibitem[Bayer et~al.(2024)]{bayer2024cysecbert}
Bayer, M., Kuehn, P., Shanehsaz, R., \& Reuter, C. (2024). CySecBERT: A Domain-Adapted Language Model for the Cybersecurity Domain. \emph{ACM Transactions on Privacy and Security}, 27(2), 1--20.

\bibitem[Jin et~al.(2023)]{jin2023darkbert}
Jin, Y., Jang, E., Cui, J., Chung, J.-W., Lee, Y., \& Shin, S. (2023). DarkBERT: A Language Model for the Dark Side of the Internet. In \textit{Proceedings of the 61st Annual Meeting of the Association for Computational Linguistics (ACL)}. arXiv:2305.08596.

\bibitem[Alam et~al.(2024)]{ctibench2024}
Alam, M.\ T., Bhusal, D., Nguyen, L., \& Rastogi, N. (2024). CTIBench: A Benchmark for Evaluating LLMs in Cyber Threat Intelligence. \textit{NeurIPS 2024 Datasets and Benchmarks Track}. arXiv:2406.07599.

\bibitem[Ruan et~al.(2024)]{vulzoo2024}
Ruan, B., Liu, J., Zhao, W., \& Liang, Z. (2024). VulZoo: A Comprehensive Vulnerability Intelligence Dataset. \textit{ASE 2024 (Demonstrations)}. arXiv:2406.16347.

\bibitem[Krippendorff(2008)]{krippendorff2008alpha}
Krippendorff, K. (2008). Systematic and random disagreement and the
reliability of nominal data. \textit{Communication Methods and Measures},
2(4), 323--338. \url{https://doi.org/10.1080/19312450802467134}.

\bibitem[Gwet(2014)]{gwet2014handbook}
Gwet, K.\ L. (2014). \textit{Handbook of Inter-Rater Reliability}.
Advanced Analytics.

\bibitem[Kapoor and Narayanan(2023)]{kapoor2023leakage}
Kapoor, S., \& Narayanan, A. (2023). Leakage and the Reproducibility
Crisis in Machine-Learning-Based Science. \textit{Patterns}, 4(9),
100804. \url{https://doi.org/10.1016/j.patter.2023.100804}.

\bibitem[Koh et~al.(2021)]{koh2021wilds}
Koh, P.\ W., Sagawa, S., Marklund, H., Xie, S.\ M., Zhang, M.,
et al. (2021). WILDS: A Benchmark of in-the-Wild Distribution Shifts.
\textit{ICML 2021}, PMLR 139.

\bibitem[Han et~al.(2017)]{severitypred2017}
Han, Z., Li, X., Xing, Z., Liu, H., \& Feng, Z. (2017).
Learning to Predict Severity of Software Vulnerability Using Only
Vulnerability Description. \textit{ICSME 2017}, 125--136.

\bibitem[Reimers and Gurevych(2019)]{sbert2019}
Reimers, N., \& Gurevych, I. (2019). Sentence-BERT:
Sentence Embeddings using Siamese BERT-Networks. \textit{EMNLP 2019}.

\bibitem[Samtani et~al.(2015)]{samtani2015assets}
Samtani, S., Chinn, R., \& Chen, H. (2015).
Exploring Hacker Assets in Underground Forums.
\textit{IEEE Intelligence and Security Informatics (ISI)}, 31--36.
\url{https://doi.org/10.1109/ISI.2015.7165935}.

\bibitem[Samtani et~al.(2017)]{samtani2017jmis}
Samtani, S., Chinn, R., Chen, H., \& Nunamaker, J.\ F. (2017).
Exploring Emerging Hacker Assets and Key Hackers for Proactive Cyber
Threat Intelligence. \textit{Journal of Management Information Systems},
34(4), 1023--1053. \url{https://doi.org/10.1080/07421222.2017.1394049}.

\bibitem[Samtani et~al.(2020)]{samtani2020dgef}
Samtani, S., Zhu, H., \& Chen, H. (2020).
Proactively Identifying Emerging Hacker Threats from the Dark Web: A
Diachronic Graph Embedding Framework. \textit{ACM Transactions on Privacy
and Security}, 23(4), Article 21. \url{https://doi.org/10.1145/3409289}.

\bibitem[Samtani et~al.(2021)]{samtani2021hap}
Samtani, S., Li, W., Benjamin, V., \& Chen, H. (2021).
Informing Cyber Threat Intelligence through Dark Web Situational
Awareness: The AZSecure Hacker Assets Portal. \textit{Digital Threats:
Research and Practice}, 2(4), Article 27.
\url{https://doi.org/10.1145/3450972}.

\bibitem[Samtani et~al.(2022)]{samtani2022eva}
Samtani, S., Chai, Y., \& Chen, H. (2022).
Linking Exploits from the Dark Web to Known Vulnerabilities for Proactive
Cyber Threat Intelligence: An Attention-Based Deep Structured Semantic
Model. \textit{MIS Quarterly}, 46(2), 911--946.

\bibitem[Zhang et~al.(2021)]{zhang2021wrench}
Zhang, J., Yu, Y., Li, Y., Wang, Y., Yang, Y., Yang, M., \&
Ratner, A. (2021). WRENCH: A Comprehensive Benchmark for Weak
Supervision. \textit{NeurIPS 2021 D\&B Track}. arXiv:2109.11377.

\bibitem[Rahman et~al.(2023)]{rahman2023cti}
Rahman, M.\ R., Mahdavi-Hezaveh, R., \& Williams, L. (2023).
What Are the Attackers Doing Now? Automating Cyberthreat Intelligence
Extraction from Text on Pace with the Changing Threat Landscape: A Survey.
\textit{ACM Computing Surveys}, 55(12), Article~241.
\url{https://doi.org/10.1145/3571726}.

\bibitem[Hughes et~al.(2024)]{hughes2024cybercrime}
Hughes, J., Pastrana, S., Hutchings, A., Afroz, S., Samtani, S.,
Li, W., \& Marin, E.\ S. (2024).
The Art of Cybercrime Community Research.
\textit{ACM Computing Surveys}, 56(6), Article~155.
\url{https://doi.org/10.1145/3639362}.

\end{thebibliography}
\end{document}